# Infrared carpet cloak designed with uniform silicon grating structure

**Xiaofei Xu, Yijun Feng[*], Yu Hao, Juming Zhao, Tian Jiang**

Department of Electronic Science and Engineering, Nanjing Univerisity, Nanjing, 210093, China

**Abstract**

Through a particularly chosen coordinate transformation, we propose an optical carpet cloak that only requires homogeneous anisotropic dielectric material. The proposed cloak could be easily imitated and realized by alternative layers of isotropic dielectrics. To demonstrate the cloaking performance, we have designed a two-dimensional version that a uniform silicon grating structure fabricated on a silicon-on-insulator wafer could work as an infrared carpet cloak. The cloak has been validated through full wave electromagnetic simulations, and the non-resonance feature also enables a broadband cloaking for wavelengths ranging from 1372 to 2000 nm.

---

[*] Author to whom correspondence should be addressed. Electronic mail: yjfeng@nju.edu.cn

Invisibility cloaking based on the transformation optics has recently intrigued lots of attention for its unprecedented ability of concealing objects from electromagnetic (EM) detection [1, 2]. Perfect invisibility cloaking could be obtained through a proper coordinate transformation which compresses a cylindrical region into a concentric cylindrical shell, leaving the electromagnetic wave detoured without interacting with any object inside the shell. After the concept has been experimentally verified in the microwave frequency [3], further efforts have been taken to apply the transformation optics to design various kinds of cloak structures and other interesting devices [4-13], for example, to simplify the cloak design and to scale the cloak from microwave to optical frequency [12, 13]. However, in most cases, the cloaks designed from the transformation optics require continuously inhomogeneous and highly anisotropic material parameters, and sometimes result in extreme or singular material properties, making the realization quite difficult.

Recently, to mitigate the material parameter constrain, a so called "carpet cloak" has been proposed [14], which could conceal an object under a curved reflecting surface with a carpet cloak by imitating the reflection of the flat surface. Such a carpet cloak could avoid both material and geometry singularities and the resulted material parameters for the cloak are in a rather moderate range, which could be realized by metamaterials with non-resonant element or conventional dielectrics. The proposed carpet cloaks have been verified and experimentally demonstrated in both microwave and optical frequency range [15-18]. These approaches all include the quasi-conformal transformation for the advantage of reducing anisotropy of the cloak. However, the material for the carpet cloak is still inhomogeneous and needs to be carefully designed to obtain the required spatial distribution of its parameters. In this letter, an alternative approach is proposed by utilizing a different coordinate transformation similar to that used to modulate the EM beam through a square block or triangular block [11, 19]. The resulted carpet cloak is made of simple homogeneous anisotropic material, which could be easily imitated and realized by alternative layers of isotropic dielectrics [12, 19-21], In a two-dimensional (2-D) configuration, we have designed an infrared carpet cloak through a uniform silicon grating structure based on a silicon-on-insulator (SOI) wafer. The cloaking performance has been verified in a wide range of infrared wavelength from 1372 nm to 2000 nm, and for different incident angles.

Although different coordinate transformations could be chosen to design invisibility cloaks, in this work we choose a straightforward one as shown in Fig. 1, in which only the *y* coordinate has been transformed in a uniform linear proportion within the cloak area. The region within the dashed right triangle in the original space (left part of Fig. 1) is compressed into another triangle region in the real space (dashed triangle in the right part of Fig. 1) leaving a small region with no coordinate meshes. Therefore under this coordinate transformation, the original space which could represents vacuum or any isotropic normal dielectric has been changed into two parts, one with twisted coordinate meshes which is equivalent to meshes inside the right triangle in the original space, representing the cloak, and the other one with no coordinate meshes, representing the concealed region, since the originally uniform EM wave propagation will be squeezed into the upper dashed triangle keeping the lower triangle unaffected by the EM waves. Such a spatial transformation could be realized with metamaterial according to the coordinate invariance of the Maxwell's equations. The coordinate transformation only affects the *y* coordinate, which could be described as $y' = (k_1 - k_2) y / k_1 + k_2 (a + x)$, where $k_1$ and $k_2$ represent the two slopes of the dashed triangle in the right of Fig. 1, respectively. The material parameters will be easily obtained following the standard procedure of transformation optics [6], which are

$$\bar{\bar{\gamma}}' = \gamma_r \begin{pmatrix} k_1/(k_1-k_2) & k_1 k_2/(k_1-k_2) & 0 \\ k_1 k_2/(k_1-k_2) & k_2^2 k_1/(k_1-k_2) + (k_1-k_2)/k_1 & 0 \\ 0 & 0 & k_1/(k_1-k_2) \end{pmatrix}, \quad (1)$$

where $\bar{\bar{\gamma}}'$ represents the permittivity tensor $\bar{\bar{\varepsilon}}'$ or permeability tensor $\bar{\bar{\mu}}'$ for the cloak in the real space, while $\gamma_r$ represents the relative permittivity $\varepsilon_r$ or permeability $\mu_r$ in the original space, respectively. It should be emphasized that the material parameters of the cloak are anisotropic but spatially invariant which could be easily realized by simple metamaterials.

For application at optical wavelength, due to the unavailable of magnetic response of natural materials, we should reduce the material parameters described by Eq. (1) to a non-magnetic form by keeping the product of $\varepsilon_x \mu_z$ or $\varepsilon_y \mu_z$ unchanged, similar to the procedure in [12]. For a transverse magnetic (TM) wave incidence (the magnetic field is perpendicular to the *x-y* plane), the material parameters are reduced to the following:

$$\bar{\bar{\varepsilon}}' = \varepsilon_r \begin{pmatrix} [k_1/(k_1-k_2)]^2 & k_2[k_1/(k_1-k_2)]^2 & 0 \\ k_2[k_1/(k_1-k_2)]^2 & k_2^2[k_1/(k_1-k_2)]^2+1 & 0 \\ 0 & 0 & 1 \end{pmatrix}, \quad \bar{\bar{\mu}}' = \begin{pmatrix} 1 & 0 & 0 \\ 0 & 1 & 0 \\ 0 & 0 & 1 \end{pmatrix}, \quad (2)$$

which represents a spatially invariant anisotropic dielectric with its optical axes rotated by a certain angle $\theta$ with the z-axis., In the local coordinate system, this anisotropic dielectric is described by a diagonal permittivity tenor $\bar{\bar{\varepsilon}}^c$, which is related to the required permittivity tensor in Eq. (2) through coordinate rotation. Therefore we have

$$\bar{\bar{\varepsilon}}' = \begin{pmatrix} \cos\theta & -\sin\theta & 0 \\ \sin\theta & \cos\theta & 0 \\ 0 & 0 & 1 \end{pmatrix} \begin{pmatrix} \varepsilon_x^c & 0 & 0 \\ 0 & \varepsilon_y^c & 0 \\ 0 & 0 & \varepsilon_z^c \end{pmatrix} \begin{pmatrix} \cos\theta & \sin\theta & 0 \\ -\sin\theta & \cos\theta & 0 \\ 0 & 0 & 1 \end{pmatrix}. \quad (3)$$

So $\bar{\bar{\varepsilon}}^c$ can be directly solved from the general procedure of eigen-value problem. For TM incidence $\varepsilon_z^c$ can be arbitrary. If we assume $\varepsilon_z^c = \varepsilon_x^c$, such a uniaxial anisotropic dielectric could be easily realized through an alternating layered system of two isotropic dielectrics [12, 19-21]. Based on the effective medium theory, if the layers are parallel to the local x-z plane, and the layer thickness is much less than the wavelength, we have the following effective permittivity elements:

$$\begin{cases} \varepsilon_z^c = \varepsilon_x^c = \dfrac{\varepsilon_1 + \eta\varepsilon_2}{1+\eta} \\ \varepsilon_y^c = \dfrac{(1+\eta)\varepsilon_1\varepsilon_2}{\eta\varepsilon_1 + \varepsilon_2} \end{cases}, \quad (4)$$

where $\varepsilon_1$ and $\varepsilon_2$ represent the permittivity of the alternating dielectric layers, and $\eta = d_2/d_1$ is the thickness ratio of the two layers, respectively.

It should be mentioned that the non-magnetic parameters introduce the boundary mismatching. The degree of mismatching could be estimated roughly by the reflectivity at the cloak boundary:

$$\Gamma = \left|\frac{Z'-Z_0}{Z'+Z_0}\right| = \left|\frac{(k_1-k_2)/k_1-1}{(k_1-k_2)/k_1+1}\right| = \left|\frac{k_2}{2k_1-k_2}\right|. \quad (5)$$

Since $k_2$ always has a non-zero value, the mismatching is inevitable. However, we could decrease the reflection from the mismatching with thick cloak to increase the ratio of $k_1$ and $k_2$.

Using the proposed triangle block described above, we can easily form a carpet cloak structure by putting together two identical triangles back to back upon a highly reflective sheet as illustrated in Fig. 2(a). Such structure can shield objects under the bumped reflective sheet from being interacted with the incoming EM waves, while the whole structure reflects the EM waves mimicking a flat mirror sheet.

To verify the concept, we propose a 2-D realizable carpet cloak structure based on a SOI wafer commonly used in silicon photonic technique. As illustrated in Fig. 2b, the silicon layer (250 nm thick) on top of a thicker silicon oxide slab (3 μm thick) layer behaves as an optical planar waveguide that confines the optical wave propagation, and TM modes propagating in the slab waveguide is similar to a 2-D wave propagation with the wave intensity only varying in the transverse directions parallel to the wafer [16, 17]. The cloak structure only requires a spatially invariant anisotropic dielectric which is realized with a uniform silicon grating structure based on the effective medium theory. In this case the alternating isotropic dielectrics are silicon ($\varepsilon_1$) and air ($\varepsilon_2$). Assuming the cloak structure is embedded in a composite background medium with a moderate permittivity of 6.55 (which can be realized by diluted silicon [16]). The whole structure is terminated at the lower end with a highly reflective boundary which could be made of either a gold layer [16] or a distributed Bragg reflector [17] consisting of alternative layers of crystal silicon and silica. The detail geometry is designed through Eq. (1) to (4). As an example to work at infrared wavelength, we choose $k_1 = 1$, $k_2 = 0.19$, $a = 2\,\mu m$, assuming an incident Gauss beam with a free-space wavelength in the range of 1372 - 2000 nm (about 536 - 781 nm in the background medium). The permittivity of silicon at such frequency is about $\varepsilon_1 = 12$. Through Eq. (3) - (4), we can get that for a grating thickness ratio of $\eta = 0.1$, the alignment angle of the grating is about $\theta = 25.5°$ and the periodicity of the silicon grating is about 110 nm.

To verify the performance, full-wave EM simulations are carried out using a commercial finite-element based EM solver (COMSOL MULTIPHYSICS©). 2D field mapping is calculated in the simulation with the four boundaries defined as perfectly matched layers (PML). Fig. 3 shows the near field magnetic field distributions for four different cases: the flat reflective boundary (mirror), the bumped reflective boundary, the anisotropic non-magnetic cloak, and the proposed silicon grating structure. A Gauss beam with a waist of 1500 nm impinges within the dielectric waveguide along the direction with an azimuth angle $\varphi = 135°$. As indicated, the magnetic field scattered by the bumped reflective boundary (Fig. 3(b)) is quite irregular, while in the case with either the anisotropic non-magnetic cloak (Fig. 3(c)), or the silicon grating structure (Fig. 3(d)) covering the bump, the magnetic scattering field is confined highly in the specular direction, mimicking that in the case of a flat reflective mirror (Fig. 3(a)).

We have also calculated the far-field pattern of the scattering field through the near-to-far-field extrapolation algorithm, and compared in Fig. 4(a) - 4(c) for the four different cases. The scattering fields are all normalized to the maximum field in the case of a flat reflective mirror. As clearly demonstrated in Fig. 4(b), when the bumped boundary is covered with either the anisotropic non-magnetic cloak or the silicon grating structure, the far-field scattering field is confined along the specular direction ($\varphi = 45°$) with little difference with that from a flat reflective boundary, while for the bare bumped reflective boundary, the scattering fields show two irregular lobes around $\varphi = 25°$ and $\varphi = 65°$. The slight backward scatterings in the cases with cloak are due to the boundary impedance mismatching. Either the near field or the far-field result shows little difference between the anisotropic non-magnetic cloak and the silicon grating structure, indicating that the approach of using alternating dielectrics to realize the anisotropic material is quite convincing. The cloak designed under the transformation optics should work for light with different incident angles. To demonstrate this, we perform more results for light incident along azimuth angle $\varphi = 90°$ and $\varphi = 150°$ in Fig. 4(a) and 4(c), respectively. In both cases, the bumped boundary with either the anisotropic non-magnetic cloak or the silicon grating structure scatters the light into a singular beam along the specular direction, although the bare bumped boundary scatters light quite differently with multiple beams.

Finally we investigate the broadband performance of the proposed carpet cloak. We have carried out simulations on the bumped boundary covered with the same silicon grating structure over a wide band of infrared wavelength from 1372 to 2000 nm. Since non-resonant element is used, the wavelength performance of the proposed carpet cloak depends mainly on the dispersion of the silicon material, which has been considered in the simulation by taking account of the material parameters at different wavelength [22]. Both near field and far-field results have been obtained, which coincide with each other. Fig. 4(d) shows the far-field scattering patterns at four different wavelengths. For wavelength range from 1372 to 2000 nm within which the silicon has a rather flattened dispersion, the whole structure scatters the infrared light mainly to the specular direction, mimicking a flat reflective boundary. In the simulations the incident Gauss beams at different wavelengths have been assumed with same waist, therefore the scattering beams tend to slightly expend with lower magnitude at large wavelength.

In conclusion, through a particular coordinate transformation, we propose an optical carpet cloak that only requires a homogeneous anisotropic dielectric material. A 2-D version of the proposed carpet cloak has been designed through the realization of the cloak with uniform silicon grating structure that could be fabricated on a SOI wafer, and the performance has been verified by full wave EM simulations of both the near field distribution and far-field scattering patterns. Due to the non-resonance of the silicon grating structure, the proposed cloak works for a wide range of infrared wavelength from 1372 to 2000 nm. Three-dimensional design could also be realized by the proposal and the carpet cloak could be constructed with alternating layers of dielectric materials.

This work is supported by the National Basic Research Program of China (2004CB719800) and the National Nature Science Foundation of China (60990322, 60671002, and 60801001).

**Figure captions**

Fig. 1. (Color online) Scheme of coordinate transformation between the original space (left) and transformed space (right). The solid lines (grey) indicate the coordinate meshes.

Fig. 2. (Color online) (a) The three-dimensional structure of the proposed carpet cloak with alternating dielectric layers. (b) 2-D version of the proposed carpet cloak designed with the silicon grating structure on a silicon-on-insulator wafer.

Fig. 3. (Color online) Magnetic field distributions for a Gaussian beam with free-space wavelength of 1696 nm impinging along the direction with an azimuth angle $\varphi = 135°$ on (a) a perfectly reflective flat boundary (mirror), (b) a bumped reflective boundary, (c) a bumped boundary covered with anisotropic non-magnetic cloak, and (d) a bumped boundary covered with silicon grating structure described in Fig. 2(b). The black line indicates the reflective boundary, and the dashed line indicates the cloak boundary.

Fig. 4. (Color online) Normalized far-field patterns of the scattering field for an incident wave along the direction with an azimuth angle of (a) 90°, (b) 135°, (c) 150°, with a free-space wavelength of 1696 nm. (d) The normalized far-field scattering patterns (135° incidence) at different wavelengths for the bumped reflective boundary covered with the silicon grating structure described in Fig. 2(b).

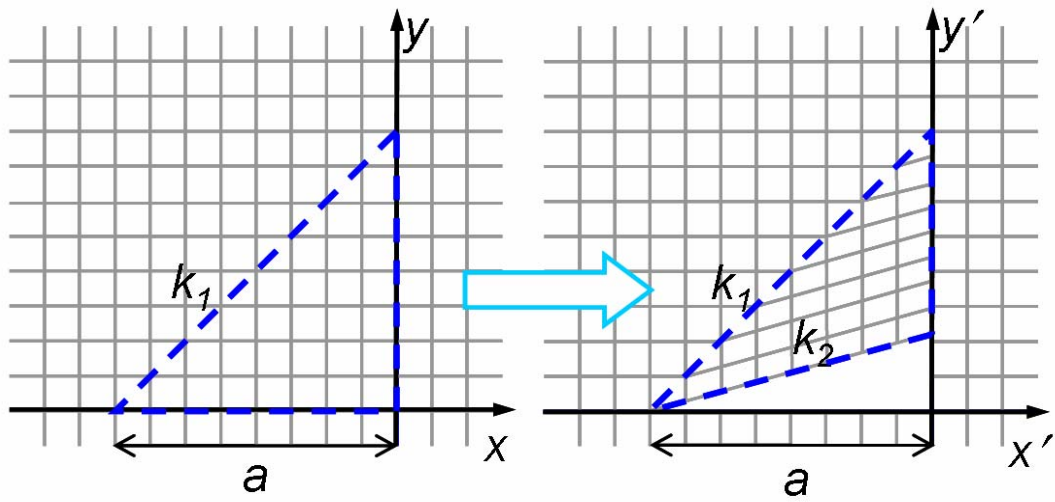

Figure 1

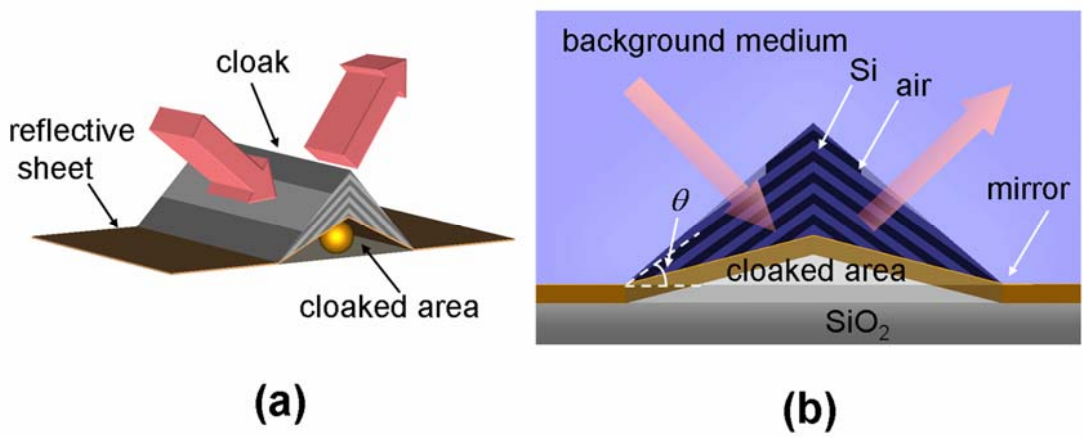

Figure 2

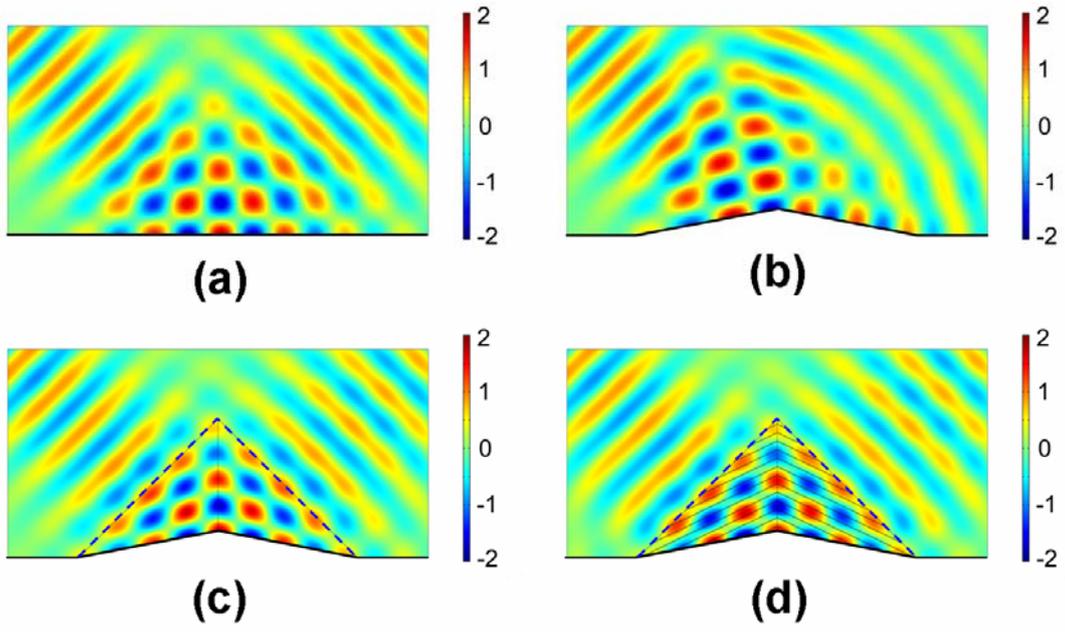

Figure 3

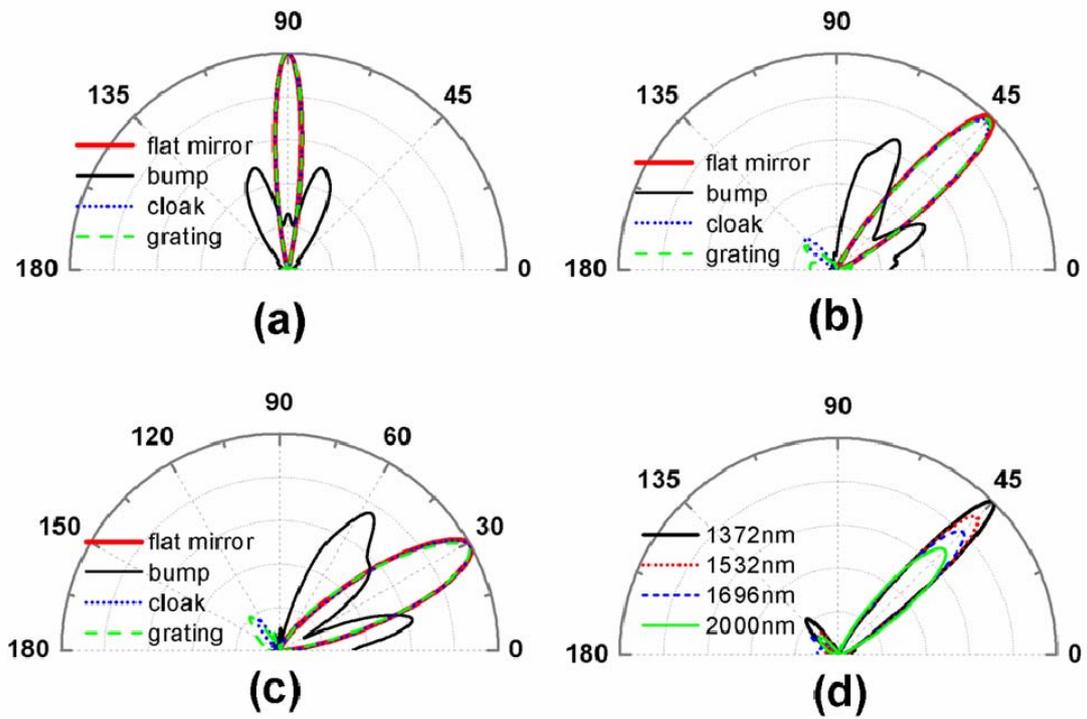

Figure 4